\documentstyle[12pt]{article}
\advance \topmargin by -\headheight
\advance \topmargin by -\headsep

\evensidemargin \oddsidemargin
\begin{document}

\topmargin -.6in

%
\def\rf#1{(\ref{eq:#1})}
\def\lab#1{\label{eq:#1}}
\def\nonu{\nonumber}
\def\br{\begin{eqnarray}}
\def\er{\end{eqnarray}}
\def\be{\begin{equation}}
\def\ee{\end{equation}}
\def\foot#1{\footnotemark\footnotetext{#1}}
\def\lb{\lbrack}
\def\rb{\rbrack}
\def\llangle{\left\langle}
\def\rrangle{\right\rangle}
\def\blangle{\Bigl\langle}
\def\brangle{\Bigr\rangle}
\def\llbrack{\left\lbrack}
\def\rrbrack{\right\rbrack}
\def\lcurl{\left\{}
\def\rcurl{\right\}}
\def\({\left(}
\def\){\right)}
\def\v{\vert}
\def\bv{\bigm\vert}
\def\Bgv{\;\Bigg\vert}
\def\bgv{\bigg\vert}
\def\dot3{\cdot\cdot\cdot}
\def\lskip{\vskip\baselineskip\vskip-\parskip\noindent}
\relax

\def\tr{\mathop{\rm tr}}
\def\Tr{\mathop{\rm Tr}}
\def\partder#1#2{{{\partial #1}\over{\partial #2}}}
\def\funcder#1#2{{{\delta #1}\over{\delta #2}}}

\def\a{\alpha}
\def\b{\beta}
\def\d{\delta}
\def\D{\Delta}
\def\eps{\epsilon}
\def\vareps{\varepsilon}
\def\g{\gamma}
\def\G{\Gamma}
\def\grad{\nabla}
\def\h{{1\over 2}}
\def\l{\lambda}
\def\L{\Lambda}
\def\m{\mu}
\def\n{\nu}
\def\o{\over}
\def\om{\omega}
\def\O{\Omega}
\def\p{\phi}
\def\P{\Phi}
\def\pa{\partial}
\def\pr{\prime}
\def\ra{\rightarrow}
\def\s{\sigma}
\def\S{\Sigma}
\def\t{\tau}
\def\th{\theta}
\def\Th{\Theta}
\def\z{\zeta}
\def\ti{\tilde}
\def\wti{\widetilde}

\def\lie{{\cal G}}
\def\dlie{{\cal G}^{\ast}}
\def\elie{{\widetilde \lie}}
\def\edlie{{\elie}^{\ast}}
\def\hlie{{\cal H}}
\def\wlie{{\widetilde \lie}}
\def\f#1#2#3 {f^{#1#2}_{#3}}                   

\def\rlx{\relax\leavevmode}
\def\inbar{\vrule height1.5ex width.4pt depth0pt}
\def\IZ{\rlx\hbox{\sf Z\kern-.4em Z}}                 
\def\IR{\rlx\hbox{\rm I\kern-.18em R}}                
\def\IC{\rlx\hbox{\,$\inbar\kern-.3em{\rm C}$}}       
\def\one{\hbox{{1}\kern-.25em\hbox{l}}}
\def\0#1{\relax\ifmmode\mathaccent"7017{#1}%
        \else\accent23#1\relax\fi}
\def\omz{\0 \omega}
%
\def\ltimes{\mathrel{\raise0.04cm\hbox{${\scriptscriptstyle |\!}$
\hskip-0.175cm}\times}}

\def\mark{\noindent{\bf Remark.}\quad}
\def\prop{\noindent{\bf Proposition.}\quad}
\def\theor{\noindent{\bf Theorem.}\quad}
\def\name{\noindent{\bf Definition.}\quad}
\def\exam{\noindent{\bf Example.}\quad}

\def\Ouc{{\cal O}_{(U_0 ,c)}}           
\def\Gsu{G_{stat}(U_0 ,c)}                   
\def\Gs{G_{stat}}                            
\def\Asu{{\lie}_{stat} (U_0 ,c)}                   
\def\As{{\lie}_{stat}}                             
\def\Suc#1{\Sigma \Bigl( #1 ; (U_0 ,c) \Bigr)}        
\def\suc#1{{\hat \sigma} (#1 ; (U_0 ,c))}
\def\sh{\hat s}                              
\def\ssh#1{{\hat \sigma}^{#1}}
\def\ruc#1{r_{#1} (U_0 ,c)}                      
\def\Y#1{Y(#1)}
\def\y{{\hat y}}
\def\yp{y_{+}(g^{-1})}
\def\YT{Y_t (g^{-1})}
\def\yt{y_\tau (g)}
\def\W1#1{W \lbrack #1 \rbrack}                
\def\Wuc#1{W \lbrack #1 ; (U_0 ,c)\rbrack}

\def\hd{{\widehat D}}
\def\dt{{\hat d}}
\def\Gpr{G^{\pr}}
\def\GT{\tilde \Gamma}
\def\GTy{\funcder {\GT} {y (t)}}
\def\GTz#1{\funcder {\GT} {y_{#1} (t)}}
\def\Ly#1{{\hat L}^{#1}_t (y)}
\def\Ry#1{R^{#1}_t (y)}
\def\LA{{\cal L}^A}

\def\Tu#1{{\widetilde \Theta}^{#1}}
\def\Td#1{{\widetilde \Theta}_{#1}}
\def\Z{\widetilde Z}
\def\T#1{{\hat {\cal T}}(#1)}
\def\dNz{\delta^{(N)} (z_1 - z_2 )}
\def\dNth{\delta^{(N)} (\th_1 - \th_2 )}
\def\dN#1#2{\delta^{(N)} ({#1} - {#2})}
\def\DN{{\llbrack D {\widetilde \Theta} \rrbrack}^2_N}
\def\Du#1{{\widetilde D}^{#1}}
\def\Dd#1{{\widetilde D}_{#1}}

\def\Tor{{\wti {\rm SDiff}}\, (T^2 )}
\def\Lh#1{{\hat {\cal L}}({#1})}
\def\M{{\cal M}}
\def\dM{{\cal M}^{\ast}}
\def\Mc{{\cal M}(R^1 \times S^1 )}
\def\dMc{{\cal M}^{\ast}(R^1 \times S^1 )}
\def\st1{\stackrel{\ast}{,}}

\def\Winf{{\bf W_\infty}}
\def\DO{DOP (S^1 )}                           
\def\DA{{\cal DOP} (S^1 )}                    
\def\eDA{{\widetilde {\cal DOP}} (S^1 )}       
\def\dDA{{\cal DOP}^{\ast} (S^1 )}                  
\def\edDA{{\widetilde {\cal DOP}}^{\ast} (S^1 )}    
\def\PsDO{\Psi{\cal DO} (S^1 )}           
\def\Rm#1#2{r(\vec{#1},\vec{#2})}          
\def\sto{\stackrel{\circ}{,}}              
\def\sta{\, ,\,}
\def\xx{(\xi , x)}
\def\yy{(\eta , y)}
\def\xxt{(\xi , x ; t )}
\def\intres{\int dx\, {\rm Res}_\xi \; }
\def\intrest{\int dt\, dx\, {\rm Res}_\xi \;}
\def\Res{{\rm Res}_\xi}
\def\pexx{e^{\pa_x \pa_\xi}}
\def\mexx{e^{-\pa_x \pa_\xi}}
\def\SLinf{SL (\infty ; \IR )}             
\def\slinf{sl (\infty ; \IR )}               
\def\sumlm{\sum_{l=1}^{\infty} \sum_{\v m\v \leq l}}
\def\WDO#1{W_{DOP (S^1 )} \lb #1\rb}               

\def\A{\cal A}
\def\AA{\widetilde {\cal A}}

\newcommand{\nit}{\noindent}
\newcommand{\ct}[1]{\cite{#1}}
\newcommand{\bi}[1]{\bibitem{#1}}

\newcommand{\NP}[1]{Nucl. \ Phys.}
\newcommand{\PL}[1]{Phys. \ Lett.}
\newcommand{\PRL}[1]{Phys.\ Rev.\ Lett. }
\newcommand{\AP}[1]{Ann.\ Phys. }
\newcommand{\IM}[1]{Inv.\ Math. }
\newcommand{\JMP}[1]{ J.\ Math.\ Phys. }
\newcommand{\RMP}[1]{Rev.\ Mod.\ Phys.}
\newcommand{\JDG}[1]{ J.\ Diff.\ Geom. }
\newcommand{\PTP}[1]{ Prog.\ Theor.\ Phys. }
\newcommand{\SPTP}[1]{Suppl.\ Prog.\ Theor.\ Phys. }
\newcommand{\PR}[1]{Phys.\ Rev. }
\newcommand{\PREP }[1]{Phys.\ Reports }
\newcommand{\NC}[1]{Nuovo \ Cim. }
\newcommand{\NCL }[1]{Nuovo\ Cim.\ Lett. }
\newcommand{\CMP}[1]{Commun.\ Math.\ Phys. }
\newcommand{\TMF}[1]{Theor.\ Math.\ Phys. }
\newcommand{\CQG}[1]{Class.\ Quant.\ Grav. }
\newcommand{\FAA}[1]{Funct. Analys. Appl. }
\newcommand{\JP}[1]{J.\ Phys. }
\newcommand{\JA }[1]{ J. Algebra }
\newcommand{\JFA}[1] {J. Funct. Anal. }
\newcommand{\MPL}[1] { Mod. Phys. Lett. }
\newcommand{\IJMP}[1] { Int. J. Mod. Phys. }
\newcommand{\RMAP}[1] { Rep.\ Math.\ Phys. }

\begin{titlepage}
\noindent\null\hfill {{\sl RI-1-92 / January, 1992 }}

\noindent\null\hfill {{\sl BGU-92 / 1 / January - PH}}

\vskip .6in

\begin{center}
{\large {\bf Induced $\, {\bf W_\infty}\,$ Gravity as a WZNW Model
}}
\end{center}

\normalsize
\vskip .4in

\begin{center}
{E. Nissimov$^{\,1}$ and
S. Pacheva\footnotemark
\footnotetext{On leave from: Institute of Nuclear Research and
Nuclear Energy, BG-1784 Sofia, Bulgaria.}}

\par \vskip .1in \noindent
Department of Physics \\
Ben-Gurion University of the Negev \\
Box 653, 84105 Beer Sheva, Israel \\
\par \vskip .3in

\end{center}

\begin{center}
I. Vaysburd

\par \vskip .1in \noindent
Racah Institute of Physics \\
Hebrew University\\
Jerusalem 91904, Israel
\par \vskip .3in
\end{center}

\begin{center}
{\large {\bf ABSTRACT}}\\
\end{center}
\par \vskip .3in \noindent

We derive the explicit form of the Wess-Zumino quantum effective action of
chiral $\, \Winf$-symmetric system of matter fields coupled to a general
chiral $\, \Winf$-gravity background. It is expressed as a geometric action
on a coadjoint orbit of the deformed group of area-preserving diffeomorphisms
on cylinder whose underlying Lie algebra is the centrally-extended algebra of
symbols of differential operators on the circle. Also, we present a systematic
derivation , in terms of symbols, of the ``hidden" $SL (\infty ; \IR )$
Kac-Moody currents and the associated $SL (\infty ; \IR )$ Sugawara form of
energy-momentum tensor component $T_{++}$ as a consequence of the
$SL (\infty ; \IR )$ stationary subgroup of the relevant $\, \Winf$
coadjoint orbit.

\end{titlepage}

{\large{\bf 1. Introduction}}
\lskip \indent
The infinite-dimensional Lie algebra $\,\Winf$ (and its generalizations
$\,{\bf W_{1+\infty}}$ etc.) \ct{Pope1,Pope2,Bakas} are nontrivial ``large $N$"
limits of the associative, but {\em non}-Lie, conformal $\,{\bf W_N}$
algebras \ct{Za85}. They arise in various problems of two-dimensional
physics. The list of their principal applications includes
self-dual gravity \ct{selfdual}, first Hamiltonian structure of
integrable KP hierarchy \ct{KP}, string field actions in the collective
field theory approach \ct{AvJev91}, conformal affine Toda theories
\ct{AFGZ}. One of the most remarkable manifestations of $\,\Winf$-type
algebras is the recent discovery of a subalgebra of their ``classical" limit
$\, {\bf w_\infty}\,$ (the algebra of area-preserving diffeomorphisms) in $c=1$
string theory as symmetry algebra of the special discrete states
\ct{KlePo91} or as the algebra of infinitesimal deformations of the
ground ring \ct{Wit91}. Also, it is worth noting that similar algebras are
found also in $D=2$ quasitopological models, such as $D=2$ Yang-Mills
\ct{qcd2},
where the metric dependence of the partition function degenerates into
a dependence on the area only.

It is well known in the mathematical literature \ct{Feig},
that the family of possible deformations $\,\Winf (q)\,$
of the initial ``classical" $\, {\bf w_\infty}\,$ depends on
a single parameter $\, q$ and that, for each fixed value of $\, q$,
$\,\Winf (q)$ possesses an one-dimensional cohomology with values in
$\IR$ . In particular, for $q=1$ one finds that  $\,\Winf (1)
\simeq \eDA\,$ - the centrally extended algebra of differential operators on
the circle, which was recently studied in refs.\ct{dop}
The equivalence of $\,\eDA\,$ to the original definition of $\,\Winf (1)
$ \ct{Pope1,Bakas} was explicitly demonstrated in \ct{BaKheKir91}.


In this letter we first derive a WZNW field-theory action
$\,\WDO{g}\,$ on a generic coadjoint orbit of the group $\,G=\DO$.
The elements $\, g\xxt\,$ of this group for fixed time $\, t\,$
are symbols of exponentiated differential operators on $S^1$ and in this
sense $\,\DO$ is the formal Lie group corresponding to the Lie algebra
$\,\eDA$ . As it was shown in \ct{shorty}, the Legendre transform
$\,\G \lb g\rb = - \W1{g^{-1}}\,$ of a group coadjoint orbit action
$\W1{g}\,$ for a general infinite-dimensional group $\, G\,$
provides the exact solution for the quantum effective action of matter
fields possessing an infinite-dimensional Noether symmetry group
$\, G_0\,$ - the ``classical" undeformed version of the group $\,G$ .
Thus, our WZNW action $\WDO{g}\,$ is the explicit field-theoretic  expression
of the induced $\,\Winf$-gravity effective action. In particular, we show that
$\,\WDO{g}\,$ reduces to the well-known Polyakov's WZNW action of
induced $D=2$ gravity in the light-cone gauge \ct{D2grav} when
restricting the WZNW field $\, g\xxt$ to the Virasoro subgroup of
$\,\DO$ . Furthermore, the appearence of the ``hidden" $\SLinf$
Kac-Moody symmetry and the associated $\SLinf\,$ Sugawara form of
the $\, T_{++}\,$ component of the energy-momentum tensor are shown to
be natural consequenses of $\, \SLinf$ stationary subgroup
the pertinent $\, \DO$ coadjoint orbit.
Also, we present WZNW field-theoretic
expressions in terms of $\, g\xxt\,$ for the ``hidden" currents and
$\, T_{++}$ .
\lskip \indent
{\large{\bf 2. Basic Ingredients}}
\lskip \indent
The object of primary interest is the infinite-dimensional
Lie algebra $\, \lie = \DA \,$ of symbols  of differential operators
\foot{Let us
recall \ct{trev} the correspondence between (pseudo)differential operators
and symbols :
$\, X \xx = \sum_k \xi^k X_k (x) \longleftrightarrow {\hat X} = \sum_k X_k (x)
{(-i \pa_x )}^k $.} on the circle $S^1$
with vanishing zero-order part $\elie =\Bigl\{\, X\equiv X \xx = \sum_{k \geq
1}
\xi^k X_k (x)\,\Bigr\}$ .
 For any pair $ X,Y \in \lie = \DA \,$ the Lie commutator is
given in terms of the associative (and  {\em non}commutative) symbol product
denoted henceforth by a cirle $\, \circ$ :
\be
\lb X \sta Y \rb  \equiv X \circ Y - Y \circ X  \;\;\; ;\;\;\;
X \circ Y \equiv X\xx \, \exp \left( \overleftarrow{\pa_\xi}
\overrightarrow{\pa_x} \right) \, Y\xx   \lab{1}
\ee
In order to determine the dual space $\, \dlie = \dDA \,$, let us consider the
space $\PsDO = \Bigl\{ U \equiv U\xx = \sum_{k=1}^{\infty} \xi^{-k}
U_k (x) \Bigr\}$ of all purely pseudodifferential symbols \ct{trev} on $S^1$
and the following bilinear form on $\PsDO \otimes \DA$ :
\be
\llangle U \v X \rrangle \equiv \intres U \circ X = \intres
\left( \, \mexx \, U\xx \, \right)\; X\xx    \lab{2}
\ee
The last equality in \rf{2} is due to the vanishing of total derivatives w.r.t.
the measure $\, \intres\,$, and $\, \Res  U\xx = U_1 (x)$ . From
\rf{2} we conclude that any pseudodifferential symbol of the form $\, U^{(0)}
= \pexx \left( {1\o {\xi}} u(x) \right)\,$ is ``orthogonal" to any
differential symbol $X \in \DA$ , i.e. $\llangle U^{(0)} \v X \rrangle = 0
\,$ . Thus, the dual space $\, \dlie = \dDA \,$ can be defined as the factor
space $\PsDO \,\backslash \,\Big\{ \pexx {1\o {\xi}} u(x)
\Bigr\}\,$ w.r.t. the ``zero" pseudodifferential symbols. In particular, we
shall adopt the definition :
\be
\dlie = \left\{ U_{\ast}\, ;\;  U_{\ast} \xx = U\xx - \pexx
\left( {1\o {\xi}} \Res U\xx \right) \; {\rm for}\; \forall U \in \Psi {\cal
DO}
\right\}  \lab{3}
\ee
Having the bilinear form \rf{2} one can define the coadjoint action of
$\lie$ on $\dlie$ via :
\br
\llangle ad^{\ast} (X) \, U \,\v\, Y \rrangle = - \llangle U\,\v\,
\lb X\sta Y \rb \rrangle  \nonu \\
\Bigl( ad^{\ast}(X)\, U\,\Bigr) \xx \equiv \lb X \sta U \rb_{\ast}
  \lab{4}
\er
Here and in what follows, the subscript $\, (-)\,$ indicates taking the part of
the symbol containing all negative powers in the $\xi$-expansion, whereas the
subscript $\,\ast\,$ indicates projecting of the symbol on the dual space
\rf{3} . The Jacobi identity for the coadjoint action $\, ad^{\ast}(\cdot )$
\rf{4} is fulfilled due to the following important property :
\be
\left\lb X\sta \pexx \left( {1\o {\xi}} u(x) \right) \right\rb_{-} =
\pexx \biggl( {1\o {\xi}} \Res \lb X \sta \pexx {{u(x)}\o {\xi}} \rb \biggr)
\lab{5}
\ee
i.e., the coadjoint action of $\DA$ on $\PsDO$ maps ``zero" pseudodifferential
symbols into ``zero" ones.

The central extension in $\elie \equiv \eDA = \DA \oplus \IR\,$ is given by the
two-cocycle $\, \omega (X,Y) =
- {1\o {4\pi}} \llangle \sh (X) \v Y \rrangle\,$, where the cocycle
operator $\, \sh : \lie \longrightarrow \dlie\, $ explicitly reads \ct{dop} :
\be
\sh (X) = \lb X \sta \ln \xi \rb_{\ast}    \lab{6}
\ee
Let us now consider the Lie group $G = \DO\,$ defined as exponentiation of the
Lie algebra $\lie\,$ of symbols of differential operators on $S^1$ :
\be
G=\left\{ g\xx = {\rm Exp} X\xx \equiv \sum_{N=0}^{\infty} {1\o {N!}} X\xx
\circ
X\xx \circ \cdot\cdot\cdot \circ X\xx \;\right\}   \lab{7}
\ee
and the group multiplication is just the symbol product
$\, g \circ h $ . The adjoint and coadjoint action of
$G =\DO$ on the Lie algebra $\DA$ and its dual space $\dDA$, respectively,
is given as :
\be
\Bigl( Ad (g) X \Bigr) = g \circ X \circ g^{-1} \;\;\;\; ;\;\;\;\;
\Bigl( Ad^{\ast} (g) U \Bigr) = \Bigl( g \circ X \circ g^{-1}
\Bigr)_{\ast}   \lab{9}
\ee
The group property of \rf{9} $\, Ad^{\ast} (g\circ h ) =
Ad^{\ast} (g) \; Ad^{\ast} (h) \,$ easily follows from the ``exponentiated"
form of the identity \rf{5}.

After these preliminaries we are ready to introduce the two interelated
fundamental objects $\, S\lb g\rb \,$ and $\, Y\lb g\rb \,$ entering
the construction of the geometric action on a coadjoint orbit of $G$.
To this end we shall follow the general formalism for geometric actions on
coadjoint orbits of arbitrary infinite-dimensional groups with central
extensions proposed in refs. \ct{ANPZ,kovhid}.
Namely, $S\lb g \rb\,$ is a nontrivial $\dlie$-valued one-cocycle
on the group $\,G\,$ (also called finite ``anomaly" or generalized Schwarzian),
whose infinitesimal form is expressed through the
Lie-algebra $\lie$ cocycle operator $\, \sh (\cdot )\,$ \rf{6}
(infinitesimal ``anomaly") :
\be
S\lb g\circ h\rb = S \lb g\rb + Ad^{\ast}(g) S \lb h\rb \;\;\;\; ;\;\;\;\;
{d \o d t} S\bigl\lb \,{\rm Exp}(t X)\,\bigr\rb \Bgv_{t=0}
= \sh (X)    \lab{10}
\ee
The explicit solution of eqs.\rf{10} reads :
\be
S \lb g\rb = - {\Bigl( \, \lb \ln \xi \sta g \rb \circ g^{-1}
\Bigr)}_{\ast}   \lab{11}
\ee
Further, $Y \lb g\rb\,$ is a $\lie$-valued one-form on the group manifold
which is related to the $\dlie$-valued group one-cocycle $S\lb g\rb\,$ via the
following basic exterior-derivative equation :
\be
dS\lb g\rb = - Ad^{\ast}(g) \, \sh \left( Y\lb g^{-1}\rb \right)   \lab{15}
\ee
The integrability condition for \rf{15} implies that the one-form $Y\lb g\rb\,$
satisfies the Maurer-Cartan equation and that it is a $\DA$-valued group
one-cocycle :
\be
dY \lb g\rb = \h \biggl\lb \, Y \lb g\rb\, \sta Y \lb g\rb\, \biggr\rb
\;\;\;\; ;\;\;\;\; Y\lb g\circ h \rb  = Y \lb g\rb + Ad (g) Y \lb h\rb
\lab{13}
\ee
{}From \rf{6} and \rf{11}-\rf{13} one easily finds :
\be
Y\lb g\rb = dg \xx \circ g^{-1}\xx    \lab{14}
\ee

At this point it would be instructive to explicitate formulas \rf{6}, \rf{11}
and \rf{14} when the elements of $G=\DO$ and $\lie =\DA$ are restricted to the
Virasoro subgroup (subalgebra, respectively) :
\br
X\xx = \xi\, \omega (x) \longleftrightarrow \omega (x) \pa_x \;\in {\cal V}ir
 \nonu  \\
g\xx = {\rm Exp}\, (\,\xi\,\omega (x)\, ) \;\;\longleftrightarrow\;\;
F(x) \equiv \exp ( \omega (x) \pa_x )\, x \;\in {\rm Diff\,}(S^1 )  \lab{16}
\er
Substituting \rf{16} into \rf{6}, \rf{11} and \rf{14}, one obtains :
\br
Y\lb g\rb \Bgv_{g\xx = {\rm Exp}\, (\,\xi\,\omega (x)\, )} = \xi {{dF (x)}\o
{\pa_x F(x)}}    \;\;\; ;\;\;\;
\sh (X) = \lb \xi \omega (x) \sta \ln \xi \rb_{\ast} = - {1\o 6} \xi^{-2}
\;\pa_x^3 \omega (x)  + \cdot\cdot\cdot   \nonu  \\
S\lb g\rb \Bgv_{g\xx = {\rm Exp}\, (\,\xi\,\omega (x)\, )} = - {1\o 6} \xi^{-2}
\; \left({{\pa_x^3 F}\o {\pa_x F}} - {3\o 2}
{\({{\pa_x^2 F}\o {\pa_x F}}\)}^2 \right) + \cdot\cdot\cdot
 \lab{17}
\er
The dots in \rf{17} indicate higher order terms $\, O(\xi^k )$, $k\geq
3$, which do not contribute in bilinear forms with elements of $\,{\cal
V}ir\,$ \rf{16}.
\lskip \indent
{\large{\bf 3. WZNW Action of $\bf W_\infty$ Gravity} }
\lskip \indent
According to the general theory of group coadjoint orbits \ct{kks}, a generic
coadjoint orbit $\Ouc$ of $G\,$ passing through a point $(U_0 ,c)$ in the
extended dual space $\edlie = \dlie \oplus \IR\,$ :
\be
\Ouc \equiv \biggl\{ \Bigl( U (g),c \Bigr) \in \edlie \; ;\;
U(g) = Ad^{\ast}(g) U_0 + c S\lb g\rb \biggr\} \lab{18}
\ee
has a structure of a phase space of an (infinite-dimensional) Hamiltonian
system. Its dynamics is governed by the following Lagrangian geometric action
written solely in terms of the interrelated fundamental group and algebra
cocycles $\, S\lb g\rb \, ,\, Y\lb g\rb\, ,\, \sh (\cdot )\, $ (cf.
eqs.\rf{6},\rf{10}-\rf{14} ) \ct{ANPZ,kovhid} :
\be
\W1{g} = \int_{\cal L} \llangle U_0 \bv Y\lb g^{-1}\rb \rrangle -
c \int \biggl\lb \Bigl\langle S\lb g\rb \bv Y\lb g\rb \Bigr\rangle - \h
d^{-1} \Bigl( \Bigl\langle \sh (Y\lb g\rb) \bv Y\lb g\rb \Bigr\rangle\Bigr)
\biggr\rb  \lab{19}
\ee
The integral in \rf{19} is over one-dimensional curve ${\cal L}$ on the phase
space $\Ouc$ \rf{18} with a ``time-evolution" parameter $\, t$ . Along the
curve ${\cal L}$ the exterior derivative becomes $\, d = dt \,\pa_t$ .
Also, $d^{-1}\,$ denotes the cohomological operator of Novikov \ct{nov} -
the inverse of the exterior derivative, defining the customary multi-valued
term present in any geometric action on a group coadjoint orbit.

In the present case of $G=\DO\,$, the co-orbit action \rf{19} takes
the following explicit form, which (as discussed in section 1) is precisely
the Wess-Zumino action for induced $\,\Winf$-gravity (the explicit dependence
of symbols on $\xxt$ will in general be suppressed below ) :
\br
W \lb g\rb = - \intrest U_0 \circ g^{-1} \circ \pa_t g  +
 \nonu   \\
{c\o {4\pi}} \int_{\cal L} \intres \Biggl( \lb \ln \xi \sta g  \rb \circ
g^{-1} \circ \pa_t g  \circ g^{-1}
-\h d^{-1} \Biggl\{ \left\lb \ln \xi \sta g^{-1} \circ
dg \right\rb \wedge \left( g^{-1} \circ dg \right) \Biggr\}\,\Biggr)   \lab{20}
\er
The physical meaning of the first term on the r.h.s. of \rf{20} is that of
coupling of the chiral $\Winf$ Wess-Zumino field $\, g=g\xxt\,$ to a
chiral $\Winf$-gravity ``background".
For simplicity, we shall consider henceforth the case $\,U_0 = 0 \,$.

It is straighforward to obtain, upon substitution of eqs.\rf{16}-\rf{17},
that the restriction of $g \xxt$ to the Virasoro subgroup reduces the $\Winf\,$
Wess-Zumino action
\rf{20} to the well-known Polyakov's Wess-Zumino action of induced $D=2\,$
gravity \ct{D2grav,AlSh89}.

The group cocycle properties (eqs.\rf{10},\rf{13} ) of $S\lb g\rb\,$ \rf{11}
and $Y\lb g\rb\,$ \rf{14} imply the following fundamental group composition law
for the $\Winf$ geometric action \rf{20} :
\be
W \lb g\circ h\rb = W \lb g\rb + W \lb h\rb - {c\o {4\pi}} \intrest \Bigl( \,
\lb \ln \xi \sta h \rb \circ h^{-1} \circ g^{-1} \circ
\pa_t g \, \Bigr)    \lab{21}
\ee
Eq.\rf{21} is a particular case for $\Winf\,$ of the group composition law for
geometric actions on coadjoint orbits of arbitrary infinite-dimensional groups
with central extensions \ct{kovhid}. It generalizes the famous
Polyakov-Wiegmann group composition law \ct{PW} for ordinary $D=2 \,$
WZNW models.

Using the general formalism for co-orbit actions in \ct{ANPZ,kovhid} we find,
that the basic Poisson brackets for $S\lb g\rb \,$ \rf{11} following from the
action \rf{20} read :
\be
\Bigl\{ \, S\lb g\rb \xx\; ,\, S\lb g\rb \yy\, \Bigr\}_{PB} =
\biggl\lb\, S\lb g\rb \xx + \ln \xi \; \sta \; \d_{DOP} (y,\eta ;x,\xi ) \,
\biggr\rb_{\ast}   \lab{pb}
\ee
where $\d_{DOP} (\cdot\, ;\cdot ) \in \dlie \otimes \lie \,$ denotes the
kernel of the $\,\d$-function on the space of differential operator symbols :
\be
\d_{DOP} (x,\xi ; y,\eta) = \pexx \left(\, \sum_{k=1}^{\infty} \xi^{-(k+1)}\,
\eta^k\, \d (x-y)\, \right)  \lab{23}
\ee
Eq.\rf{pb} is a succinct expression of the Poisson-bracket realization of
$\Winf\,$, which becomes manifest by rewriting \rf{pb} in the equivalent form :
\be
\Bigl\{ \, \llangle S\lb g\rb \v X\rrangle\; ,\,\llangle S\lb g\rb \v Y\rrangle
\, \Bigr\}_{PB} = - \llangle S\lb g\rb \v \,\lb X \sta Y \rb \rrangle +
\llangle \sh (X) \v Y \rrangle       \lab{pb1}
\ee
for arbitrary fixed $X,Y \in \lie=\DA\,$. Alternatively, substituting into
\rf{pb} (or \rf{pb1}) the $\xi$-expansion of the
pseudodifferential symbol $S\lb g\rb \xx = \sum_{r\geq 2} \xi^{-r} S_r (x)\,$,
one recovers the Poisson-bracket commutation relations for $\,\Winf\;$
among the component fields $\,S_r (x)\,$ in the basis of ref.\ct{BaKheKir91}
(which is a ``rotation" of the more customary $\,\Winf\;$ basis of
refs.\ct{Pope2} ).

In particular, for the component field $S_2 (x) \equiv {{4\pi}\o c}
T_{--}(x)\,$ (the energy-momentum tensor component, cf. \rf{17})
one gets from \rf{pb} the Poisson-bracket realization of the Virasoro algebra :
\be
\left\{ S_2 (x)\, ,\, S_2 (y) \right\}_{PB} = - {{4\pi}\o c}
\Bigl(\, 2 S_2 (x) \pa_x \d (x-y) + \pa_x S_2 (x)\, \d (x-y)
+ {1\o 6} \pa_x^3 \d (x-y) \, \Bigr)     \lab{pb2}
\ee
The higher component fields $S_r (x)\;,\, r=3,4,\cdot\cdot\cdot$ turn out to be
quasi-primary conformal fields of spin $\,r\,$. The genuine primary fields
${\cal W}_r (x) \;\, (r\geq 3)\,$ are obtained from $S_r (x)\,$ by adding
derivatives of the lower spin fields $S_q (x)\;\, (2\leq q\leq r-1)\,$. For
instance, for ${\cal W}_3 (x) = S_3 (x) - {3\o 2}\pa_x S_2 (x)\,$, eq.\rf{pb}
yields :
\be
\left\{ S_2 (x)\, ,\, {\cal W}_3 (y) \right\}_{PB} =  - {{4\pi}\o c} \Bigl( \,
3 {\cal W}_3 (x) \pa_x \d (x-y) + 2\pa_x {\cal W}_3 (x)\, \d (x-y) \, \Bigr)
\lab{pb3}
\ee
\lskip\indent
{\large{\bf 4. Noether and ``Hidden" Symmetries of $\bf W_\infty$ Gravity} }
\lskip \indent
The general group composition law \rf{21} contains the whole information
about the symmetries of the $\Winf$ geometric action \rf{20}. First, let us
consider arbitrary infinitesimal left group translation .
The corresponding variation of the action \rf{20} is
straighforwardly obtained from \rf{21} :
\be
\d_\vareps^L \W1{g} \equiv \W1{(\one + \vareps )\circ g} - \W1{g} =
{c\o {4\pi}} \intrest \biggl\{\,\Bigl(\,\lb \ln \xi \sta g \rb \circ
g^{-1} \Bigr)_{\ast} \circ \pa_t \vareps \,\biggr\}   \lab{24}
\ee
{}From \rf{24} one finds that \rf{20} is invariant under $\,t$-independent left
group translations and the associated Noether conserved current is the
generalized ``Schwarzian" $S\lb g\rb \,$ \rf{11} whose components are the
(quasi)primary conformal fields $S_r (x;t)\,$ of spin $\, r\,$ .

Next, let us consider arbitrary right group translation .
Now, from \rf{21} the variation of the $\Winf$ action \rf{20} is given by :
\br
\d_\z^R \W1{g} \equiv \W1{g\circ (\one + \z )} - \W1{g} = -
{c\o {4\pi}} \intrest \Bigl(\, \lb \ln \xi \sta \,\z \rb_{\ast} \circ \YT
\Bigr) =  \lab{26a}  \\
{c\o {4\pi}} \intrest \Bigl(\,\lb \ln \xi \sta \YT \rb_{\ast}
\,\circ \,\z \Bigr)  \lab{26b}
\er
where $\, \YT\,$ denotes the Maurer-Cartan gauge field :
\be
\YT = - g^{-1} \circ \pa_t g          \lab{MC}
\ee
Equality \rf{26b} implies the equations of motion \foot{The restriction of
eq.\rf{27} to the Virasoro subgroup via \rf{16}-\rf{17} takes the well known
from \ct{D2grav} $\;\pa_x^3 \, \left( \,\pa_t f \backslash \pa_x f \,\right) =
0 \;$, and $f(x;t)\,$
is the inverse Virasoro group element : $f(\,F(x;t)\,;\,t\,)=x\,$.} :
\be
\sh \left( \, \YT\, \right)\Bgv_{\rm on-shell} = 0    \lab{27}
\ee
As a matter of fact, the {\em off}-shell relation \rf{15} exhibits the full
equivalence between the Noether conservation law $\, \pa_t S\lb g\rb = 0\,$
\rf{24} and the equations of motion \rf{27}.

On the other hand, equality \rf{26a} shows that the $\Winf$ geometric action
\rf{20} is {\em gauge}-invariant under arbitrary {\em time-dependent}
infinitesimal right-group translations $\; g\xxt \longrightarrow g\xxt \circ
(1 + {\wti \z}\xxt )\,$ which satisfy :
\be
\sh ({\wti \z}) \equiv - \left\lb \ln \xi \sta \,{\wti \z} \right\rb_{\ast}
= 0      \lab{28}
\ee
For finite right group translations $\, k= {\rm Exp}\, {\wti \z}\,$
the integrated form of \rf{28} reads :
\be
S\lb k\rb \equiv - \Bigl(\,\lb \ln \xi \sta k\rb \circ
k^{-1}\Bigr)_{\ast} =  0    \lab{29}
\ee
The solutions of eqs. \rf{28} and \rf{29} form a
subalgebra in $\DA$, and a subgroup in $\DO$ , respectively. From \rf{18} one
immediately concludes that the latter subgroup :
\be
\Gs = \Bigl\{\, k\; ;\; S\lb k\rb = 0 \Bigr\}       \lab{30}
\ee
is precisely the stationary subgroup of the underlying coadjoint
orbit $\, {\cal O}_{(U_0 = 0 \, ,\, c)}$ . The Lie algebra of \rf{30} :
\be
\As = \left\{\, {\wti \z}\; ;\; \sh ({\wti \z}) \equiv
- \left\lb \ln \xi \sta \,{\wti \z} \right\rb_{\ast} = 0 \;\right\} \lab{31}
\ee
is the maximal centerless (``anomaly-free") subalgebra of $\eDA$ , on which the
cocycle \rf{6} vanishes :
$\,\omega ({\wti \z}_1 ,{\wti \z}_2 )
= - \llangle \sh ({\wti \z}_1 )\,\v\,{\wti \z}_2 \,\rrangle = 0\;$ for any pair
${\wti \z}_{1,2} \in \As\,$.

The full set of linearly independent solutions $\,\Bigl\{\,\z^{(l,m)}\xx \,
\Bigr\}\,$ of $\, \sh ({\wti \z}) =0\,$, comprising a basis in
$\,\As\,$ \rf{31}, can be written in the form :
\be
\z^{(l,m)} \xx = \sum_{q=1}^l {l\choose q} {{(l-1)! (l+q)!}\o {(q-1)! (2l)!}}\,
{{\xi^q \,x^{q+m}}\o {\G (q+m+1)}}      \lab{33}
\ee
where $\, l=1,2,\cdot\cdot\cdot \;$ and $\, m=-l,-l+1,\cdot\cdot\cdot
,l-1,l\,$.

The basis \rf{33} identifies the stationary sublagebra $\,\As\,$ \rf{31} as the
infinite-dimensional algebra $\,\slinf$. Namely, $\, \As$ decomposes (as a
vector space) into a direct sum of irreducible representations $\,
{\cal V}^{(l)}_{sl (2)}\,$ of its $\, sl (2;\IR )$ subalgebra with
spin $\, l\,$ and unit multiplicity :
$ \,\As = \bigoplus_{l=1}^{\infty} \, {\cal V}^{(l)}_{sl (2)}$ .
This $\, sl(2;\IR )\;$ subalgebra is generated by the symbols
$\, 2\z^{(1,1)} = \xi x^2 \; ,\; \z^{(1,0)} = \xi x\,$ and
$\, \z^{(1,-1)} = \xi\,$ .
The subspaces $\,{\cal V}^{(l)}_{sl (2)}\,$ are spanned by the
symbols $\, \Bigl\{\,\z^{(l,m)}\; ;\; l= {\rm fixed}\; ,\;
\v m \v \leq l \,\Bigr\}\,$ with $\,\z^{(l,l)}\,$ being the highest-weight
vectors  :
\be
\lb \xi x^2 \sta \z^{(l,l)} \rb = 0 \;\; ;\;\;
\lb \xi x \sta \z^{(l,m)} \rb = m \,\z^{(l,m)}  \;\; ;\;\;
\lb \xi \sta \z^{(l,m)} \rb = \z^{(l,m-1)}    \lab{sl2}
\ee
The Cartan subalgebra of $\,\slinf$ is spanned by the subset $\, \Bigl\{\,
\z^{(l,0)}\; ,\, l=1,2,\cdot\cdot\cdot \Bigr\}\,$ of symbols \rf{33}.

The above representation of $\,\slinf\,$ in terms of symbols \rf{33} is
analogous to the construction of $\,\slinf\,$ as ``wedge" subalgebra
$\, W_{\wedge} (\mu )\,$ of $\, \Winf\,$ for $\, \mu =0 $ \ct{Pope2,slinf},
which in turn is isomorphic to the algebra $\, A_{\infty}\,$ of Kac \ct{Kac85}.

Now, accounting for \rf{31}-\rf{33}, one can write down explicitly the solution
to the equations of motion \rf{27} :
\be
\YT\Bgv_{\rm on-shell} =
\sumlm J^{(l,m)}(t) \,\z^{(l,m)}\xx       \lab{34}
\ee
with $\,\z^{(l,m)} \,$ as in \rf{33}.
The coefficients $\, J^{(l,m)}(t)\,$ in \rf{34} are arbitrary functions of
$\, t\,$ and represent the on-shell form of the currents of the ``hidden"
$\, \As \equiv \slinf\,$ Kac-Moody symmetry of $\W1{g}\,$ \rf{20}.

Indeed, upon right group translation with $\, \z_{(a)} = \sum
a^{(l,m)}(x,t) \circ \z^{(l,m)}\xx\,$ with arbitrary coefficient functions
(zero order symbols) $\,a^{(l,m)}(x,t) \,$, one obtains from \rf{26a} :
\br
\d_{\z_{(a)}}^R \W1{g} = - {c\o {4\pi}} \int dt\, dx \, \sumlm \pa_x
a^{(l,m)}(x,t) \; {\wti J}^{(l,m)}(x,t)
 \lab{35}   \\
{\wti J}^{(l,m)} \equiv \sum_{r=1}^{\infty} {(-\pa_x )}^{r-1} \left\{
{(-1)^l} l! (l+1)! {{x^m}\o {\G (m)}} \Bigl(\,\YT\,\Bigr)_r
- {1\o {r+1}} \pa_x \Bigl(\, \z^{(l,m)} \circ \YT \Bigr)_r \right\}   \lab{36}
\er
The subscripts $\, r\;$ in \rf{36} and below indicate taking the coefficient
in front of $\, \xi^r\,$ in the corresponding symbol.

The Noether theorem implies from \rf{35} that $\,{\wti J}^{(l,m)}(x,t) \,$
\rf{36}
are the relevant Noether currents corresponding to the symmetry of the $\Winf$
action \rf{20} under arbitrary right group $\SLinf$ translations.
Clearly, $\,{\wti J}^{(l,m)}(x,t) \,$ are $\,\slinf$-valued and are conserved
w.r.t. the ``time-evolution" parameter $\, x \equiv x^{-}$ :
\be
\pa_x \,{\wti J}^{(l,m)}(x,t) \;\Bgv_{\rm on-shell} = 0      \lab{37}
\ee
Substituting the on-shell expression \rf{34} into \rf{36} we get :
\be
{\wti J}^{(l,m)}(x,t) \;\Bgv_{\rm on-shell} = \sumlm K^{(l,m)
(l^{\pr},m^{\pr})}
 J^{(l^{\pr},m^{\pr})} (t)
\lab{38}
\ee
where $\, K^{(l,m) (l^{\pr},m^{\pr})}\,$ is a constant invariant symmetric
$\slinf$ tensor :
\be
K^{(l,m) (l^{\pr},m^{\pr})} =
\sum_{r=1}^{\infty} {(-\pa_x )}^{r-1} \left\{ \,\Bigl(\,
\z^{(l^{\pr},m^{\pr})}\,\Bigr)_r \, \Res \Bigl\lb \ln \xi \sta \z^{(l,m)}
\Bigr\rb - {1\o {r+1}}\pa_x \Bigl(\, \z^{(l,m)} \circ \z^{(l^{\pr},m^{\pr})}
\Bigr)_r \right\}
\lab{39}
\ee
naturally representing the Killing metric of $\slinf$ .

The fact, that the currents $\, J^{(l,m)} (t)\,$ in \rf{34} generate a
$\slinf$ Kac-Moody algebra, can be shown most easily by considering
infinitesimal right group translation $\, g \longrightarrow g \circ (\one +
\z_\vareps )\,$ with $\, \z_\vareps = \sum_{l,m} {\vareps}^{(l,m)}(t)\,
\z^{(l,m)}\xx \in \slinf\;$ on $\, \YT \equiv - g^{-1} \circ \pa_t g\,$ .
Recall (cf. \rf{35},\rf{38} ), that $\, J^{(l,m)} (t)\,$ are the corresponding
Noether symmetry currents. From the cocycle property \rf{13} one obtains :
\be
\d_{\z_\vareps}^R \YT \equiv
Y_t \left( (\one - \z_\vareps )\circ g^{-1} \right) - \YT
= -\pa_t \z_\vareps + \left\lb \YT \sta \z_\vareps \right\rb   \lab{40}
\ee
which upon substitution of \rf{34} yields :
\be
\d_\vareps J^{(l,m)} (t) = - \pa_t \vareps^{(l,m)}(t) +
\f{(l,m)}{(l^{\pr},m^{\pr})}{(l^{\pr\pr},m^{\pr\pr})} J^{(l^{\pr},m^{\pr})} (t)
\vareps^{(l^{\pr\pr},m^{\pr\pr})}(t)  \lab{41}
\ee
Here $\,\f{(l,m)}{(l^{\pr},m^{\pr})}{(l^{\pr\pr},m^{\pr\pr})}\, $
denote the the structure constants of $\slinf$
in the basis $\,\z^{(l,m)}\,$ \rf{33} ( i.e., $\lb \z^{(l,m)}\,\sta\,
\z^{(l^{\pr},m^{\pr})}\rb  =
\f{(l,m)}{(l^{\pr},m^{\pr})}{(l^{\pr\pr},m^{\pr\pr})}
\z^{(l^{\pr\pr},m^{\pr\pr})}\,$ ).

Finally, let us also show that the canonical Noether energy-momemtum tensor
$\, T_{++}\,$ (the Noether current corresponding to the symmetry of the
$\Winf$ action \rf{20} under arbitrary rescaling of $\, t\equiv x^{+}$ )
authomatically has the (classical) Sugawara form in terms of the ``hidden"
$\slinf$ Kac-Moody currents $\, J^{(l,m)} (t)$ \rf{34}. Indeed, the variation
of
\rf{20} under a reparametrization $\, t \longrightarrow t + \rho (t,x)\,$
reads :
\br
\d_\rho \W1{g} = - {1\o {4\pi}} \int dt\, dx\, \pa_x \rho (t,x)\; T_{++}
(t,x)      \lab{42}    \\
T_{++} \equiv {1\o {2c}} \sum_{r=0}^{\infty} {{(-1)^r}\o {r+1}} \pa_x^r
\left\{ \,\Bigl(\,\YT \circ \YT\,\Bigr)_r +
{1\o {r!}} \Res \left( \pa_\xi^{r+1} \YT\,\circ
\, \Bigl\lb \ln \xi \sta \YT \Bigr\rb \;\right) \,\right\}   \lab{43}
\er
Substituting \rf{34} into \rf{43} and accounting for \rf{33},
one easily gets the $\,\slinf$ Sugawara
representation of the energy-momentum tensor \rf{43} :
\be
T_{++} (t,x) \Bgv_{\rm on-shell} =  {1\o {2c}} \sum_{(l,m) ,
(l^{\pr},m^{\pr})}
K^{(l,m)(l^{\pr},m^{\pr})} J^{(l,m)} (t)\, J^{(l^{\pr},m^{\pr})} (t)  \lab{44}
\ee
where $\, K^{(l,m)(l^{\pr},m^{\pr})}\,$ is the $\slinf$ Killing metric tensor
\rf{39}.

In particular, substituting into \rf{43} the restriction of $g\xxt$ to the
Virasoro subgroup via \rf{16}-\rf{17}, we recover the well-known (classical)
$sl (2;\IR )$ Sugawara form of $\, T_{++}$ in $D=2$ induced gravity
\ct{Po88}.
\lskip \indent
{\large{\bf 5. Conclusions and Outlook} }
\lskip \indent
According to the general discussion in \ct{shorty}, the Legendre transform
$\,\G \lb y\rb = - \W1{g^{-1}}\,$ of the induced $\,\Winf$-gravity WZNW action
\rf{20} is the generating functional, when considered as a functional of
$\, y \equiv \YT\,$, of the quantum correlation functions of generalized
Schwarzians $\, S\lb g\rb $ .
Similarly, $\,W \lb J\rb \equiv - \WDO{g}\,$, when considered
as a functional of
$\, J \equiv  - {c\o {4\pi}} S\lb g\rb\,$, is the generating functional of all
correlation functions of the currents $\,\YT $ .
These correlation functions can be straightforwardly obtained,
recursively in $\, N\,$, from the functional differential equations
(i.e., Ward identities) :
\be
\pa_t \funcder{\G}{y} + \left\lb \funcder{\G}{y} - {c\o {4\pi}} \ln \xi \sta
y\, \right\rb_{\ast} = 0   \;\;\; ;\;\;\;
\pa_t J + \left\lb \funcder{W}{J} \sta J  - {c\o {4\pi}} \ln \xi
\right\rb_{\ast} = 0       \lab{ward}
\ee
An interesting problem is to derive the $\; \Winf\,$ analogue of the
Knizhnik-Zamolodchikov equations \ct{KniZa84} for the correlation functions
$\, \llangle g (\xi_1 ,x_1 ;t_1 ) \dot3 g (\xi_N ,x_N ;t_N ) \rrangle\, $ .
To this end we need the explicit form of the symbol
$\, r\Bigl( \xx ; (\xi^{\pr} , x^{\pr}) \Bigr) \; \in \DA \otimes \DA\,$
of the classical $r$-matrix of
$\,\Winf\,$ . This issue will be dealt with in a forthcoming paper.

Another basic mathematical problem is the study of the complete classification
of the coadjoint orbits of $\, \DO\,$ and the classification of its highest
weight irreducible representations.

Let us note that, in order to obtain the WZNW action of induced
$\, W_{1+\infty}\,$
gravity along the lines of the present approach, one should start with the
algebra of differential operator symbols containing a nontrivial zero order
term in the $\,\xi$-expansion $\; X=X_0 (x) + \sum_{k \geq 1} \xi^k X_k (x)\,$.
In this case one can solve the ``hidden" symmetry (i.e. the ``anomaly" free
subalgebra) equation $\; \left\lb \ln \xi \sta \z \, \right\rb_{(-)} = 0\;$ and
the result is the Borel subalgebra of $\, gl (\infty ;\IR )\;$ spanned by the
symbols $\, \z^{(p,q)} = \xi^p x^q \,$ with $\, p \geq q\,$ . The
$\, W_{1+\infty}\,$ WZNW action will have formally the same form as \rf{20},
however, now the meaning of the symbol $\, g^{-1} \xxt\,$ of the inverse group
element is obscure due to the nontrivial ($\xx$-dependent) zero order
term in the $\,\xi$-expansion of $\, g\xxt$ .
\lskip \indent
{\bf Acknowledgements.} It is a pleasure to thank S. Elitzur, V. Kac,
D. Kazhdan, A. Schwimmer and A. Zamolodchikov for useful comments and
illuminating discussion.

\end{document}